\documentclass[twocolumn,showpacs,preprintnumbers,amsmath,amssymb]{revtex4}
\usepackage{graphicx}
\usepackage{dcolumn}
\usepackage{bm}

\newcommand{\ie}{{\it i.e.}}

\newcommand{\tr}{\rm tr}

\begin{document}

\title{Spin relaxation of two-dimensional electrons with a hierarchy of spin-orbit couplings}

\author{Yuan Li and You-Quan Li}
\affiliation{Zhejiang Institute of Modern Physics and Department
of Physics,\\ Zhejiang University, Hangzhou 310027, P. R. China}

\begin{abstract}
The density matrix formalism is applied to calculate the
spin-relaxation time for two-dimensional systems with a hierarchy
of spin-orbit couplings, such as Rashba-type, Dresselhaus-type and
so on. It is found that the spin-relaxation time can be
infinite if those coupling strengths $\alpha$, $\beta$, $\gamma_1$
and $\gamma_2$ satisfy either condition (i) $\alpha=\beta, ~\gamma_1=0$
or (ii) $\alpha=-\beta,~\gamma_2=0$, which correspond to the vanishing Yang-Mills
magnetic field. The effect caused by the application
of an external magnetic field is also discussed.
It is found that the longitudinal and in-plane spin components can
possess infinite life time when the spin components, the Larmor
precession frequency and the external magnetic field satisfy
certain relations.
\end{abstract}

\received{\today} 

\maketitle

\section{Introduction }

Spintronics~\cite{Zutic}, or spin based electronics~\cite{Wolf},
has absorbed much more attention during the last decade. One
important issue in this research area is the manipulation of spin
polarized electrons with the help of an electric
field~\cite{Datta,Schliemann,Rashba2003,Levitov,Murakami}. The
system with spin-orbit couplings makes these efforts possible and
thus brings great interests from both academic and practical
aspects~\cite{Datta,Schliemann,Rashba2003,Levitov}.
However, the problem of the loss of the average microscopic spin is
crucial in experimental data analysis and
applicable device construction.
The study of the spin-relaxation mechanisms of two-dimensional electrons
is thus very important.

Spin relaxation exhibits some properties of the spin dynamics,
which plays inevitable role in realizing the applicable
spintronics devices.
The main mechanism of spin relaxation in systems
lacking inversion symmetry is the D'yakonov-Perel's
mechanism~\cite{Dyakonov1971,Dyakonov1986}, in which the spin of
the electron precesses due to an effective $\mathbf{k}$-dependent magnetic field.
For electrons in two-dimensional
semiconductor heterostructures or quantum wells, the structure
inversion asymmetry brings about the Rashba spin-orbit
coupling~\cite{Rashba1960,Bychkov-jetp,Bychkov-jpc}, while the
bulk inversion asymmetry in the $A_3B_5$ compounds leads to the
Dresselhaus~\cite{Dresselhaus}  spin-orbit coupling.
Spin-relaxation time in some semiconductors with both Rashba and
Dresselhaus couplings was calculated by analyzing the condition of
spin decay~\cite{Averkiev1999,Averkiev2002}, and the effect of
external magnetic fields was discussed~\cite{Glazov2004}
furthermore. An infinite spin-relaxation time~\cite{Zhang2006} was
predicted in the system with equal Rashba and Dresselhaus
coupling constants by making use of an SU(2) symmetry in $k$-space.
It is important to understand the spin-relaxation
mechanism and the condition for infinite spin-relaxation time to
occur, which would be helpful for overcoming the difficulties in
the spin-based information processes.

In this paper, we develop the aforementioned theory of spin
relaxation to describe two-dimensional electron systems in the
presence of $U(1)$ Maxwell field and $SU(2)$ Yang-Mills fields.
Such a system can be realized in certain semiconductor materials
where the spin-orbit couplings, such as Rashba-type,
Dresselhaus-type and etc., play crucial roles. Using density
matrix formalism, we calculate the spin-relaxation time for the
system with a hierarchy of spin-orbit couplings. In the absence of
the Maxwell magnetic field, the infinity of the spin-relaxation
time occurs if the spin-orbit couplings $\alpha$, $\beta$,
$\gamma_1$ and $\gamma_2$ satisfy the condition in which the
Yang-Mills magnetic field vanishes.
In order to capture the physical essence of the emergence of an infinite spin-relaxation
time, we further study the effect of the external magnetic field on the
same systems and find that the longitudinal and in-plane spin
components can also possess infinite life times when the spin
orientation, the Larmor precession frequency and external magnetic
field satisfy some relations.
Base on the analysis of spin-orbit systems with or without Maxwell magnetic fields,
we expose a physics picture for a clear understanding of the
infinite spin-relaxation time, which is helpful for the design of
the spin-based devices.

\section{Spin relaxation arising from spin-orbit couplings}

To start with a general formalism, we consider the Schr\"{o}dinger
equation for a particle moving in an external $U(1)$ Maxwell field
and an $SU(2)$  Yang-Mills field~\cite{Jin},
\begin{eqnarray}
  &&i\hbar\frac{\partial}{\partial t}\Psi(r,t)=H\Psi(r,t), \nonumber \\
  && H= \frac{1}{2m}(\textbf{\^{p}}-\frac{e}{c}\textbf{A}
  -\eta\mathcal{A}^a \hat{\tau}^a )^2+eA_0+\eta \mathcal{A}^a_0\hat{\tau}^a,
\end{eqnarray}
where $\Psi$ is a two-component wavefunction,
$\textbf{A}_{\mu}=(A_0,A_i)$ denotes the vector potential of
the Maxwell electromagnetic field, and
$\mathbb{A}_{\mu}=\displaystyle\mathcal{A}_\mu^a \hat{\tau}^a$
that of Yang-Mills field with $\hat{\tau}^a$ being the generators
of $SU(2)$ Lie group. It has been shown~\cite{Jin} that the
Yang-Mills fields can be realized in certain semiconductor
materials.

Firstly, we consider a two-dimensional system (in $x$-$y$ plane)
with a four-parameter Yang-Mills potentials:
$ \mathcal{\vec{A}}_0=(0, 0, 0)$,
$\mathcal{\vec{A}}_x=\frac{2m}{\eta \hbar}(0, \beta+\alpha, \gamma_2)$,
$\mathcal{\vec{A}}_y=\frac{2m}{\eta \hbar}(\beta-\alpha, 0, \gamma_1)$,
$\mathcal{\vec{A}}_z=(0,\, 0,\, 0)$
where $\alpha$, $\beta$ and $\gamma$ characterize the strengths
of spin-orbit couplings of Rashba-type, Dresselhaus-type, etc., respectively.
If writing out the Hamiltonian explicitly, we have
\begin{eqnarray}\label{eq:hami}
H &=& \frac{\hbar^2k^2}{2m}+V+k_y\sigma_x(\alpha-\beta)-k_x\sigma_y(\alpha+\beta)
     \nonumber \\
  &&-(\gamma_1 k_y+\gamma_2k_x)\sigma_z
     \nonumber \\
 &=&\frac{\hbar^2k^2}{2m}+V
  +\frac{\hbar}{2}\vec{\sigma}\cdot\mathbf{\Omega_k},
\end{eqnarray}
with
$V=2m(\gamma_1^2+2(\alpha^2+\beta^2)+\gamma_2^2)/\hbar^2+V_{sc}$ ,
$\mathbf{\Omega_k}=2(k_y(\alpha-\beta),-k_x(\alpha+\beta),
-(\gamma_1k_y+\gamma_2k_x))/\hbar$ and $\vec{\sigma}$ being the
pauli matrix. Here $m$ stands for the effective mass of electrons in
the material, $V_{\rm sc}=e A_0$ the scattering potential which is
independent of spin indices. The scattering is supposed to be
elastic. The last term in above equation
$H'=\frac{\hbar}{2}\vec{\sigma}\cdot\mathbf{\Omega_k}$ causes the
precession of electron spins with the Larmor frequency
$\mathbf{\Omega_k}$ which can be regarded as an effective magnetic
field.

To calculate the spin-relaxation time, we use density matrix
formalism. The electron density matrix $\rho(\mathbf{k})$ with
components $\rho_{ss'}(\mathbf{k})$, $s,s'$ being the indices of
electron spin states, is defined by~\cite{Averkiev1999,Pikus}
\begin{eqnarray}\label{eq:md1}
\frac{\rho(\mathbf{k})}{\tau}+\frac{i}{\hbar}[H'(\mathbf{k}),\rho(\mathbf{k})]+
\sum_{k'}W_{\mathbf{kk'}}(\rho(\mathbf{k})-\rho(\mathbf{k'}))=0,
\end{eqnarray}
where $\tau$ is the lifetime, $W_{\mathbf{kk'}}$ is the scattering
probability from $\mathbf{k}$ to $\mathbf{k}'$ and the square
bracket denotes commutator.

Since $H'$ contributes merely a small perturbation, the
spin-relaxation time is much longer than the time for
electron-momentum distribution to become isotropic, \ie,
$\tau\gg\tau_1$, $\tau_1$ the momentum relaxation time. Therefore,
it is convenient to split the density matrix into two parts,
\begin{eqnarray}
\rho=\overline{\rho} + \rho', \quad\textrm{with}\quad \overline{\rho'}=0.
\end{eqnarray}
Here we use a bar to denote an average taken over all directions
of \textbf{k} and a prime to denote the deviation ones,
$\rho'(\mathbf{k})\ll\overline{\rho}$. Taking average for
Eq.~(\ref{eq:md1}), we have the following relation,
\begin{eqnarray}\label{eq:md2}
\frac{\overline{\rho}}{\tau}
 +\frac{i}{\hbar}\overline{[H'(\mathbf{k}),\rho'(\mathbf{k})]}=0.
\end{eqnarray}
Eq.~(\ref{eq:md1}) can also be written out as
\begin{eqnarray}\label{eq:md3}
\frac{\rho'(\mathbf{k})}{\tau}
 +\frac{i}{\hbar}[H'(\mathbf{k}),\rho'(\mathbf{k})]
 -\frac{i}{\hbar}\overline{ [H'(\mathbf{k}),\rho'(\mathbf{k})]}
 +\frac{i}{\hbar}[H'(\mathbf{k}),\overline{\rho}]
    \nonumber\\
 +\sum_{\mathbf{k'}}W_{\mathbf{kk'}}
[\rho'(\mathbf{k})-\rho'(\mathbf{k'})]=0, \hspace{26mm}
\end{eqnarray}
in which Eq.~(\ref{eq:md2}) has been used.
Without taking account of the higher order terms,
we need to solve the following equation
\begin{eqnarray}\label{eq:md3}
\frac{i}{\hbar}[H'(\mathbf{k}),\overline{\rho}]+\sum_{\mathbf{k'}}W_{\mathbf{kk'}}
[\rho'(\mathbf{k})-\rho'(\mathbf{k'})]=0.
\end{eqnarray}
This approximation is valid when $\Omega_k\tau_1 \ll 1$. For
elastic process the scattering probability is a function of
deflection angle only, which makes it possible to expand the above
equation in terms of Fourier series. After some algebra, one can
express $\rho'$ in terms of $\overline\rho$. Substituting it into
Eq.~(\ref{eq:md2}) and employing the Boltzmann equation with only
collision term, we can obtain the rate of averaged density matrix
$\overline\rho$, namely,
\begin{eqnarray}\label{eq:md4}
(\frac{\partial \overline{\rho}}{\partial
t})_{\rm sp.rel.}=-\frac{1}{\hbar^2}\sum_n
\tau_n[H'_{-n},~ [H'_n,\overline{\rho}]~],
\end{eqnarray}
with
\begin{eqnarray}
H'_n &=& \oint
\frac{d\phi_\mathbf{k}}{2\pi}H'(\mathbf{k})\exp(-in\phi_\mathbf{k}),
  \\
\label{eq:momen}
\frac{1}{\tau_n}&=&\oint d\theta W_{\mathbf{kk'}}(1-\cos n\theta),
\end{eqnarray}
where $\phi_\mathbf{k}$ is the angle between $\textbf{k}$ and
$x$-axis, and $\theta=\phi_\mathbf{k}-\phi_{\mathbf{k'}}$.

Now we are in the position to investigate the kinetics of the spin
density $S_i(t)=\int a^{}_i (\varepsilon, t) d\varepsilon$, where
$a_i(\varepsilon, t)=[F_+ (\varepsilon) - F_-(\varepsilon) ]
s_i(t)$, $s_i(t)= \tr(\sigma_i \overline{\rho})$ and $F_\pm
(\varepsilon)$ refers to the distribution function projected along
the direction parallel or anti-parallel to $\textbf{s}= (s_1, s_2,
s_3)$. Accordingly, we obtain the evolution equation for the spin
density at the time longer than $\tau_1$~\cite{Averkiev2002}:
\begin{eqnarray}\label{eq:evolution}\displaystyle
&& \dot{S}_i(t)=-\frac{1}{\tau_{ij}}S_j(t)
   \nonumber \\[2mm]
&& \frac{1}{\tau_{ij}}=\frac{1}{2\hbar^2}\sum^{\infty}_{-\infty}
 \frac{\displaystyle\int d\varepsilon (F_+ -F_- )
       \tau_n \tr\bigl\{[ H'_{-n},\,[H'_n,\,\overline{\rho}]\,]\sigma_i\bigr\} }
  {\displaystyle\int d\varepsilon (F_+ -F_- )},
  \nonumber\\
\end{eqnarray}
where $i, j =x, y, z$.

For the Hamiltonian under consideration (\ref{eq:hami}), we obtain
the following,
\begin{eqnarray}\label{eq:tensor}
&&\frac{1}{\tau_{xx}}=\frac{\gamma_1^2+\gamma_2^2+(\alpha+\beta)^2}{2}\Lambda,
  \nonumber\\
&&\frac{1}{\tau_{yy}}=\frac{\gamma_1^2+\gamma_2^2+(\alpha-\beta)^2}{2}\Lambda,
     \nonumber\\
&&\frac{1}{\tau_{zz}}=(\alpha^2+\beta^2)\Lambda,
     \nonumber\\
&&\frac{1}{\tau_{xz}}=\frac{1}{\tau_{zx}}=\frac{\gamma_1(\alpha-\beta)}{2}\Lambda,
     \nonumber\\
&&\frac{1}{\tau_{yz}}=\frac{1}{\tau_{zy}}=-\frac{\gamma_2(\alpha+\beta)}{2}\Lambda,
     \nonumber\\
&&\frac{1}{\tau_{xy}}=\frac{1}{\tau_{yx}}=0,
\end{eqnarray}
where the coefficient $\Lambda$ is given by
\begin{eqnarray*}
\Lambda=\frac{8m}{\hbar^4}
  \frac{\displaystyle\int d\varepsilon[F_+(\varepsilon)-F_-(\varepsilon)]\tau_1(\varepsilon) \varepsilon}
  {\displaystyle\int d\varepsilon [F_+(\varepsilon)-F_-(\varepsilon)]}.
\end{eqnarray*}
The above entities in
Eq.~(\ref{eq:tensor})
define the spin-relaxation
tensor
$\Gamma=\mathrm{mat}(\frac{1}{\tau_{ij}})$.
Diagonalizing this matrix (see Appendix \ref{sec:absence} for details), we obtain
\begin{eqnarray}\label{eq:Tinverse}
{\mathbb T}^{-1}=\frac{\Lambda}{2}\left( {\begin{array}{*{100}c}
   \frac{1}{\tau^{}_{\perp}} & 0 & 0 \\[3mm]
   0 & \frac{1}{\tau^{}_{\parallel,\pm}} & 0 \\[3mm]
   0 &0 & \frac{1}{\tau^{}_{\parallel,\pm}}  \\
\end{array}} \right),
\end{eqnarray}
with
\begin{eqnarray}\label{eq:benzhen}
\frac{1}{\tau^{}_{\parallel,\pm}} &=& \frac{1}{2}
  \left\{\frac{1}{\tau^{}_{\perp}}
  \pm\sqrt{(\gamma_1^2+\gamma_2^2)^2+8\alpha\beta(\gamma_2^2-\gamma_1^2+2\alpha\beta)}
    \right\}
        \nonumber\\[2mm]
\frac{1}{\tau^{}_{\perp}} &=& \gamma_1^2+\gamma_2^2+2(\alpha^2+\beta^2)
\end{eqnarray}
Clearly, two of the diagonal elements are always positively
definite and the other one $1/\tau_{\parallel,-}$ is not. The
condition for a vanishing $1/\tau_{\parallel,-}$ turns out to be
\begin{eqnarray}\label{eq:condition}
\gamma_1^2(\alpha + \beta)^2+\gamma_2^2(\alpha-\beta)^2
  +(\alpha + \beta)^2(\alpha-\beta)^2 = 0.
\end{eqnarray}
The above equation gives rise to two solutions
\begin{eqnarray}\label{eq:qiugen}
& {\rm (i)} &   \alpha=\beta,\quad  \gamma_1=0,
    \nonumber \\
& {\rm (ii)}&  \alpha=-\beta, \,\, \gamma_2=0.
\end{eqnarray}
Under these conditions, ${\mathbb T}^{-1}_{yy}$ and ${\mathbb
T}^{-1}_{zz}$ are zero when
the infinite spin-relaxation times emerge.

Actually, the Yang-Mills ``magnetic'' field $\mathbb{B}_z=b^a_{}
\hat{\tau}^a$ can be calculated, namely
\begin{eqnarray}
b_1 =\frac{4m^2}{\eta^2\hbar^2} (\beta+\alpha)\gamma_1,
  \nonumber\\
b_2 =\frac{4m^2}{\eta^2\hbar^2} (\beta-\alpha)\gamma_2,
  \nonumber\\
b_3 =\frac{4m^2}{\eta^2\hbar^2} (\alpha^2-\beta^2).
\end{eqnarray}
The condition~(\ref{eq:condition}) is also equivalent to
\[
|~\vec{b}~|^2 =0, \quad \vec{b} =(b_1, b_2, b_3),
\]
which implies that the module of the Yang-Mills ``magnetic'' field
vanishes. In other words, the spin-relaxation time can be infinite
when the module of the Yang-Mills magnetic field is null.
This result is expected to be a criterion to evaluate whether
there has an infinite spin-relaxation time in two-dimensional
systems with spin-orbit couplings.
In order to determine which spin component has an infinite life time,
the spin precession needs to be analyzed concretely.

The nonvanishing spin orbit coupling $\gamma_1$ or $\gamma_2$ will
bring about some new features which may be useful for possible
design with an infinite spin relaxation time. For the first case
$\alpha=\beta$ and $\gamma_1=0$ in Eq.~(\ref{eq:qiugen}), the
Hamiltonian $H'$ reduces to
\begin{eqnarray}\label{eq:weiyao}
H'=\frac{\hbar}{2}\vec{\sigma}\cdot\mathbf{\Omega_k} =-k_x(2\alpha
\sigma_y+\gamma_2\sigma_z)
\end{eqnarray}
The orientation of  Larmor precession frequency
$\mathbf{\Omega_k}$ is parallel to $y'$-axis as illustrated in
Fig.~\ref{fig:nofield} in appendix A, thus one can understand why
the life time of the $S_{y'}$ component is infinite while the
other two decay. Here the $y'$-axis is defined in the
diagonalization procedure of the spin-relaxation time tensor
Eq.(\ref{eq:tensor}) for the first case given in appendix
\ref{sec:absence}.

From the appendix~\ref{sec:absence} and the figure therein,
we can see that the strengths of
spin-orbit coupling $\gamma_2$ and $\alpha$ determine the angle
$\theta'$ between $\mathbf{\Omega_k}$ and $z$-axis.
The angle $\theta'$ can be manipulated by these two parameters,
thus a definite alignment of spin with infinite life
time can be realized with the help of tuning spin-orbit
coupling strengths. On the other hand, the ratio of
different type of spin-orbit coupling constants can be determined by
means of measuring spin-relaxation time experimentally.

For the second case in Eq.~(\ref{eq:qiugen}), similar analysis can
be carried out, which is omitted here. In the special case when both
$\gamma_1$ and $\gamma_2$ vanish, one component of the tensor of
spin-relaxation time becomes null, \ie, $1/\tau_{xx}=0$ or
$1/\tau_{yy}=0$, and thus the $ S_x$ or $S_y$ has infinite life
time, which is just the case considered in Ref.~\cite{Averkiev1999}.

\section{Spin relaxation affected by external magnetic field}

In previous section, we considered the case of
$\vec{\mathcal A}_0 =0$ which means external magnetic
field is absent. In the presence of the magnetic field,
we should take account of
\begin{eqnarray}
\vec{\mathcal A}_0 =-\frac{2\mu_B}{\eta}(B_x, B_y, B_z)
\end{eqnarray}
where $\mu_B$ is the Bohr magneton.
The existence of an external magnetic field is known to
affect the dynamics of the electron's spin. The Larmor precession
of electron's spin around  a sufficiently strong longitudinal
magnetic field will suppress the precession about the internal
random magnetic fields~\cite{Dyakonov1973}. The cyclotron motion
will change the wave vector $\mathbf{k}$ and affect the spin
relaxation due to the DP mechanism. The density-matrix
formalism is applicable for calculating  electrons' spin-relaxation
time. It will be convenient to expand the density matrix for
electrons in terms of the unit and Pauli matrices,
\begin{eqnarray}
\rho(\mathbf{k})=f_\mathbf{k}+\mathbf{s}^{}_{\mathbf k}\cdot
\vec{\sigma}
\end{eqnarray}
where $f_\mathbf{k}=tr[\rho(\mathbf{k})/2]$ is the
spin-averaged-electron-distribution function and
$\mathbf{s}^{}_{\mathbf k}=tr[\rho(\mathbf{k})\vec{\sigma}]$ is the
spin per $\mathbf{k}$-state electron. The kinetic equation for the
spin distribution is given by~\cite{Glazov2003,Ivchenko,Glazov2004}:
\begin{eqnarray}\label{eq:kine}
\frac{\partial \mathbf{s}^{}_{\mathbf k}}{\partial t}
  & + & \mathbf{s}^{}_{\mathbf k}\times (\vec{\omega}^{}_L
 +\mathbf{\Omega_k})+\vec{\omega}_C\cdot[\mathbf{k}
 \times
\nabla^{}_{\mathbf k}\mathbf{s}^{}_{\mathbf k} ]
 \nonumber\\
 &+&\sum_{\mathbf{k}'}W_{\mathbf{kk'}}(\mathbf{s_k-s_{k'}})=0,
\end{eqnarray}
in which the second term refers to spin precession caused by
spin-orbit couplings given in Eq.~(\ref{eq:hami})
together with the external magnetic field;
the third term is related to the wave vector variations due to the cyclotron
motion whose frequency is $\omega_C=\frac{eB_z}{mc}$, and the last term denotes
the collision integral.
Since the internal random magnetic field is regarded as a
perturbation, \ie, $\mathbf{\Omega_k}\tau_1\ll 1$,
we can split the spin distribution function $\mathbf{s_k}$ in the following,
\begin{eqnarray}\label{eq:sk}
\mathbf{s}^{}_{\mathbf k}=\mathbf{s}^0_k + \delta \mathbf{s}^{}_{\mathbf k}
\end{eqnarray}
where $\mathbf{s}^0_k$ is a quasi-equilibrium distribution
function and thus is independent of direction of $\mathbf{k}$.
Whereas, $\delta \mathbf{s}^{}_{\bf k}$ is a nonequilibrium correction
arising from spin-orbit couplings as well as other internal
random magnetic fields and thus it contains only first angular
harmonics of the spin distribution~\cite{AAA} because elastic scattering processes
is taken into account only, accordingly
\begin{eqnarray}\label{eq:corr}
\delta
\mathbf{s_k}=\mathbf{R}_1\cos(\phi_k) + \mathbf{R}_2\sin(\phi_k),
\end{eqnarray}
where the two vectors $\mathbf{R}_1 $ and $\mathbf{R}_2$ are
irrelevant to the direction of the wave vector $\mathbf{k}$
though they are functions of the module of $\mathbf{k}$ in general.
Substituting Eq.~(\ref{eq:sk}) and (\ref{eq:corr}) into Eq.~(\ref{eq:kine}),
we obtain the following equations:
\begin{eqnarray}
\label{eq:cor1}\frac{d\mathbf{s}^0_k}{dt}
  &+&\mathbf{s}^0_k \times
 \vec{\omega}_L+\delta \mathbf{s_k}\times \mathbf{\Omega_k}=0,
   \\    \label{eq:cor2}
\frac{d\delta\mathbf{s}^{}_k}{dt}&+&\mathbf{s}^0_k \times
 \mathbf{\Omega_k}+\delta \mathbf{s_k}\times\vec{\omega}_L
   \nonumber \\
 &+&\vec{\omega}_C\cdot [\mathbf{k}\times\nabla_{\mathbf{k}}\mathbf{\delta
s_k}]+\frac{\mathbf{\delta s_k}}{\tau_1}=0.
\end{eqnarray}
where $\tau_1$ is momentum-relaxation time
whose definition is also given by Eq.~(\ref{eq:momen}) for $n=1$.
In the light of the number of total electrons  $N=2\sum_{\mathbf{k}}f_k$
and the single electron spin
$\mathbf{S}^0=\displaystyle\frac{\sum_\mathbf{k}\mathbf{s}^{}_{\mathbf k} }{N}$,
summing Eq.~(\ref{eq:cor1}) over the wave vectors,
we obtain the balance equation describing electron's spin relaxation
\begin{eqnarray}
\frac{d\mathbf{S}^0}{dt} + \mathbf{S}^0 \times \vec{\omega}_L
  + \hat{\Gamma}\mathbf{S}^0 =0,
\end{eqnarray}
where the spin-relaxation tensor $\hat{\Gamma}$
referring to the inverse of spin-relaxation times
is defined as
\begin{eqnarray}
\hat{\Gamma}\mathbf{S}^0 = \frac{1}{N} \sum_\mathbf{k}
  \delta\mathbf{s}^{}_k\times \mathbf{\Omega}_{\mathbf k}.
\end{eqnarray}
The nonequilibrium correction $\delta\mathbf{s}^{}_{\mathbf k}$ can be
obtained from Eq.~(\ref{eq:cor2}) in which the contribution of the
rate $d\delta \mathbf{s_k}/dt$ is negligible because
its magnitude is of higher order in $\mathbf{\Omega_k}\tau_1$.

\begin{figure}[h]
\includegraphics[width=4.4cm]{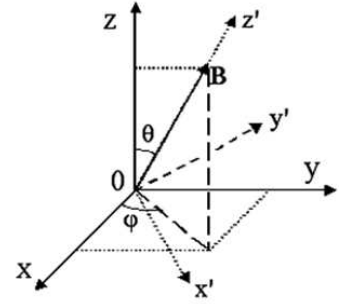}
\caption{The scheme of coordinate frames. $z$-axis parallel with
the $[001]$ growth axis. $\theta$ and $\varphi$ are the polar and
the azimuthal angles of the external magnetic field $\mathbf{B}$.
$z'$ is chosen in alignment with the orientation of $\mathbf{B}$,
$y'$ is lying in $x$-$y$ plane, and $x'$ is chosen to form a
right-hand triple with $y'$ and $z'$. }
\label{fig:rotate}
\end{figure}

Firstly, we rotate the original coordinate $\{\hat x, \hat y, \hat z
\}$ which is related to the principal crystal axes to the new one
$\{\hat x', \hat y', \hat z' \}$ (illustrated in
Fig.~\ref{fig:rotate}). The coordinates in both systems are
related, $(\hat{x}, \hat{y}, \hat{z})= (\hat{x}', \hat{y}',
\hat{z}')R^T$, by
\begin{eqnarray}\label{eq:yaozhen1}
R= \left( {\begin{array}{*{100}c}
   \cos\theta \cos\varphi, & -\sin\varphi, & \sin\theta\cos\varphi \\[1mm]
   \cos\theta \sin\varphi, &  \cos\varphi, & \sin\theta\sin\varphi \\[1mm]
   -\sin\theta,            &      0,       & \cos\theta  \\
\end{array}} \right)
\end{eqnarray}
Here, $\theta$ is the angle between $z$ and $z'$, and $\varphi$
the angle between $y'$ and $y([010])$. Similar equations are valid
for momentum components, $k_i=R_{ij}k'_{j}$ (here $i,j=x,y,z$). It
is convenient to calculate the nonequilibrium correction
$\delta\mathbf{s}_{\mathbf k}$ and the components of the
spin-relaxation tensor in the new frame of coordinate where the
Larmor frequency vector $\vec{\omega}_L = \frac{\mu_B}{\hbar}B_j
\hat{e}_j$ in the original coordinate becomes $\vec{\omega}_L
=\omega_L\hat{z'}$  in the new coordinate.

After tedious calculation, we obtain the spin-relaxation tensor
(inverse of the spin-relaxation time) $\mathbf{\hat{\Gamma}}$ for
degenerate electrons with Fermi energy $E_F$ which is given in
appendix \ref{sec:presence}. These results are valid for arbitrary
random internal magnetic field and arbitrary orientation of the
external field, from which we obtain several conclusions that will
be illustrated respectively.
\begin{figure}
\includegraphics[width=4.4cm]{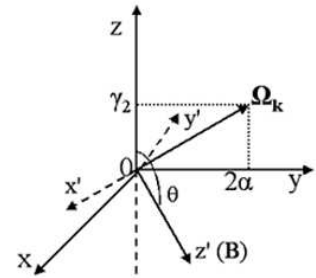}
\renewcommand{\figurename}{Fig.}
\caption{\label{fig:plane} Schematic of the case
$\Gamma_{x'x'}=0$. $\alpha=\beta$, $\gamma_1=0$, $\varphi=\pi/2$,
$\theta=\tan^{-1}(-\gamma_2/2\alpha)$, the external magnetic field
$\mathbf{B}$ is parallel to $z'$-direction, and $x'$ axis is
antiparallel to the Larmor frequency $\mathbf{\Omega_k}$.}
\end{figure}

\subsection{Longitudinal relaxation}

The longitudinal spin-relaxation rate is $1/\tau^{}_L=\Gamma_{z'z'}$.
From Eq.~(\ref{eq:rate1}), we can obtain the following conclusion:
\noindent$\Gamma_{z'z'}=0$ when either
\begin{eqnarray*}
 \alpha=\beta, \gamma_1=0, \varphi=\pi/2,
 \; \theta=\tan^{-1}(2\alpha/\gamma_2),
\end{eqnarray*}
or
\begin{eqnarray}\label{eq:case12}
\alpha=-\beta, \gamma_2=0, \varphi=0, \;
\theta=-\tan^{-1}(2\alpha/\gamma_1).
\end{eqnarray}
When the longitudinal spin-relaxation time $\tau^{}_L$ is
infinite, the spin component $S_{z'}$ has an infinite life time.
Certainly, the Hamiltonian $H'$ describing the electron spin
precession arising from spin-orbit coupling can also be written as
Eq.~(\ref{eq:weiyao}) for the former case in
Eq.~(\ref{eq:case12}). One can see from Fig.~(\ref{fig:nofield})
that $S_{z'}$ is the component parallel to $\mathbf{\Omega_k}$
when $\theta'=\theta=\tan^{-1}(2\alpha/\gamma_2)$ and
$\varphi=\pi/2$. Thus the infinite life time of $S_{z'}$ can
easily be understood from physical point of view, that is to say,
$S_{z'}$ will not precess about Larmor precession frequency
$\mathbf{\Omega_k}$ and external magnetic $\mathbf{B}$ when
$\mathbf{\Omega_k}$ is parallel with $\mathbf{B}$. While the other
components $S_{x'}$ and $S_{y'}$ have finite life time due to the
precession around $\mathbf{\Omega_k}$ arising from internal random
magnetic field. If both of the strength of spin-orbit couplings
$\gamma_1$ and $\gamma_2$ vanish, the longitudinal component of the
tensor $\Gamma_{z'z'}$ is zero when $\theta=\pi/2,~\varphi=\pi/2$,
which can be seen from the Eq.~(\ref{eq:rate1}). This recovers the
special case discussed in Ref.~\cite{Glazov2004}.

For the latter case in Eq.~(\ref{eq:case12}), the Hamiltonian $H'$
becomes
\begin{eqnarray}\label{eq:weiyao1}
H'=\frac{\hbar}{2}\vec{\sigma}\cdot\mathbf{\Omega_k} =k_y(2\alpha
\sigma_x-\gamma_1\sigma_z).
\end{eqnarray}
The Larmor frequency
$\mathbf{\Omega_k}=\displaystyle \frac{2k_y}{\hbar}
  ( 2\alpha,\, 0,\, -\gamma_1 )$
parallels to the external magnetic field $\mathbf{B}
(\theta=-\tan^{-1}(2\alpha/\gamma_1),\varphi=0 )$. So $S_{z'}$
does not decay because $\mathbf{\Omega_k}$  parallels to
$\mathbf{B}$.

\subsection{Transverse relaxation}

Let us analyze the spin relaxation in the plane perpendicular to
the external magnetic field.
One can  find that the transverse components of spin-relaxation
tensor can also be zero
(\ie, $\Gamma_{x'x'}=0$, $\Gamma_{x'y'}=0$, $\Gamma_{y'x'}=0$) when either
\[
\alpha=\beta, \gamma_1=0, \varphi=\pi/2,
\theta=-\tan^{-1}(\frac{\gamma_2}{2\alpha}),
\]
or
\begin{equation}\label{eq:case34}
\alpha=-\beta, \gamma_2=0, \varphi=0,
\theta=\tan^{-1}(\frac{\gamma_1}{2\alpha}).
\end{equation}
As shown in Fig.~\ref{fig:plane}, the $x'$ axis is antiparallel to
the Larmor frequency (\ie, $\hat{x}'\parallel -\mathbf{\Omega_k}$)
for the former case in Eq.~(\ref{eq:case34}). The spin component
$S_{x'}$  does not precess about the Larmor precession frequency
$\mathbf{\Omega_k}$. We know that $\mathbf{B}$ is perpendicular to
$\mathbf{\Omega_k}$ from Fig.~\ref{fig:plane}, thus $S_{x'}$ will
precess about the constant external magnetic field $\mathbf{B}$ in
the plane paralleling with $\mathbf{\Omega_k}$. So the random
internal magnetic fields and external magnetic field can not
induce spin relaxation for  the spin component $S_{x'}$. Then
$S_{x'}$ has an infinite life time associating with
$\Gamma_{x'x'}=0$. And the admixture of the $x'$ component to $y'$
component of spin-relaxation tensor (which is described by
$\Gamma_{x'y'}$ , $\Gamma_{y'x'}$) are zero, namely
$\Gamma_{x'y'}=0$ and $\Gamma_{y'x'}=0$ which can also be
calculated from Eqs.~(\ref{eq:rate1}-\ref{eq:rate5}).

For the latter case in Eq.(\ref{eq:case34}), the direction of the
external magnetic field is also perpendicular to the Larmor
frequency $\mathbf{\Omega_k}$ as illustrated in
Eq.~(\ref{eq:weiyao1}). The spin component $S_{x'}$  is
antiparallel to the Larmor frequency ($\hat{x}'\parallel
-\mathbf{\Omega_k}$). The random internal magnetic field and
external magnetic field will not induce spin relaxation for the
spin component $S_{x'}$ due to the same reason as aforementioned.
Then $S_{x'}$ has an infinite life time, and the components of
spin-relaxation tensor $\Gamma_{x'y'}$, $\Gamma_{y'x'}$ vanish
(see Eqs.~(\ref{eq:rate1}-\ref{eq:rate5}).

The components $\Gamma_{x'z'}$ and $\Gamma_{y'z'}$ are smaller
than others when the external magnetic field is sufficiently
strong ($\Omega^2_k\tau_1\ll \mathbf{\omega_L}$). Under this
condition the in-plane spin components rapidly rotate and the
admixture of the in plane components to $z'$-component (which is
described by $\Gamma_{x'z'}$ , $\Gamma_{y'z'}$) plays no role in
the spin dynamics. Therefore the above result manifests the
general solutions of spin-relaxation time for the Hamiltonian
Eq.~(\ref{eq:hami}).

\section{Summary}

In the above, we developed a consistent theory of spin dynamics to
describe particles moving in an external $U(1)$ Maxwell field and
an $SU(2)$ Yang-Mills field which characterizes spin-orbit
couplings in certain semiconductors (such as Rashba-type,
Dresselhaus-type or other complex types). We used the
density-matrix formalism to calculate the spin-relaxation time in
such systems in the absence and in the presence of the external
magnetic field, respectively. In the absence of external magnetic
field, we find that the spin component $S_{y'}$ or $S_{z'}$  has
an infinite life time if the strengths of spin orbit couplings
$\alpha$, $\beta$, $\gamma_1$ and $\gamma_2$ satisfy either (i)
$\alpha=\beta,~ \gamma_1=0$ or (ii) $\alpha=-\beta,~ \gamma_2=0$.
In such a case, the Yang-Mills magnetic field vanishes. From these
conditions, the direction of the spin component with infinite life
time can be manipulated by tuning $\alpha$, $\gamma_2$ or $\beta$,
$\gamma_1$ respectively. In the presence of the external magnetic
field, we considered the magnetic effect on the two-dimensional
system. We obtained that the longitudinal spin-relaxation time is
infinite when the $S_{z'}$ is parallel to $\mathbf{\Omega_k}$ and
$\mathbf{B}$ if either  (i') $\alpha=\beta$, $\gamma_1=0$,
$\varphi=\pi/2$, $\theta=\tan^{-1}(2\alpha/\gamma_2)$ or (ii')
$\alpha=-\beta$, $\gamma_2=0$, $\varphi=0$,
$\theta=-\tan^{-1}(2\alpha/\gamma_1)$. By making use of analysis
in detail, we conclude that the in-plane spin component $S_{x'}$
can also have infinite life time if $S_{x'}$ is antiparallel with
$\mathbf{\Omega_k}$ and perpendicular to $\mathbf{B}$ for either
(iii') $\alpha=\beta$, $\gamma_1=0$, $\varphi=\pi/2$,
$\theta=\tan^{-1}(-\gamma_2/2\alpha)$ or (iv') $\alpha=-\beta$,
$\gamma_2=0$, $\varphi=0$, $\theta=\tan^{-1}(\gamma_1/2\alpha)$.
That is to say, electron spins have infinite life times if they
precess in the plane parallel to the vector of the Larmor
precession frequency arising from spin-orbit couplings. These
solutions provide a better understanding on the spin dynamics of
two-dimensional system with a four-parameter Yang-Mills
potentials, which characterizes a hierarchy of spin-orbit coupling
in certain semiconductor materials. It is expected to expose some
more clues for manipulating spin via certain spin-orbit couplings
in semiconductors or elaborating spintronics storage devices with
long spin-relaxation time.

\section*{Acknowledgement}

We acknowledge helpful communications with M. M. Glazov. This work is
supported by NSFC No. 10225419 and No. 10674117.

\appendix

\section{On the diagonalizing bases}\label{sec:absence}

We can easily obtain the matrix $U$
\begin{eqnarray}
U=\displaystyle\left( {\begin{array}{*{100}c}
  1 & 0 & 0 \\[3mm]
   0 &\displaystyle\frac{2\alpha}{\sqrt{\gamma_2^2+4\alpha^2}} &
   \displaystyle\frac{-\gamma_2}{\sqrt{\gamma_2^2+4\alpha^2}} \\[3mm]
   0 &\displaystyle\frac{\gamma_2}{\sqrt{\gamma_2^2+4\alpha^2}} &
   \displaystyle\frac{2\alpha}{\sqrt{\gamma_2^2+4\alpha^2}}  \\
\end{array}} \right)
\end{eqnarray}
which diagonalizes the spin-relaxation tensor given by Eq.~(\ref{eq:tensor})
to be Eq.~(\ref{eq:Tinverse}),
\begin{eqnarray}\label{eq:hamiweiyao}
\mathbb{T}^{-1}=U^{-1}\Gamma~U
\end{eqnarray}
This matrix $U$ turns the evolution equation for
the spin density Eq.(\ref{eq:evolution}) to be,
\begin{eqnarray}\label{eq:hamiweiyao}
\frac{dS'}{dt} &=& U^{-1}\frac{dS}{dt}=-(U^{-1}\Gamma\,U) U^{-1}S
    \nonumber\\
&=&-\mathbb{T}^{-1}S'
\end{eqnarray}
\begin{figure}[h]
\includegraphics[width=4.4cm]{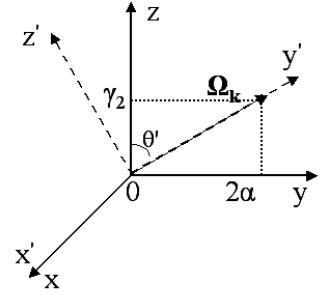}
\renewcommand{\figurename}{Fig.}
\caption{The orientation of the Larmor frequency
$\mathbf{\Omega_k}$ for the case $\alpha=\beta$, $\gamma_1=0$,
which defines the $y'$-axis of the new frame of coordinate}
\label{fig:nofield}
\end{figure}
Hence, the new spin components $S'=U^{-1}S$ is obtained
\begin{eqnarray*}
S'_{x} & = & S_x , \\
S'_{y} & = & \sin\theta'S_y+\cos\theta'S_z, \\
S'_{z} & = & -\cos\theta'S_y+\sin\theta'S_z,
\end{eqnarray*}
where $\tan\theta'=2\alpha/\gamma^{}_2$.
This means the existence of  spin-orbit coupling $\gamma_2$
makes the orientation of the Larmor frequency
$\mathbf{\Omega_k}$ to change from $y$ to $y'$-axis.
As illustrated in Fig.~\ref{fig:nofield},
the $\theta'$ refers to the angle between the Larmor frequency and $z$-axis.

\section{The spin-relaxation tensor}\label{sec:presence}

In the presence of magnetic field, we obtained the following
spin-relaxation tensor
\begin{widetext}
\begin{eqnarray}
\Gamma_{z'z'}&=&
 \frac{2k^2\tau_1}{\hbar^2D_+D_-}\{[1+\tau_1^2 (\omega_c^2+\omega_L^2)]
  \times[(\alpha+\beta)^2\cos^2\varphi+(\alpha-\beta)^2\sin^2\varphi
    \nonumber \\
& & + (\gamma_2\sin\theta-(\alpha+\beta)\cos\theta\sin\varphi)^2
  +(\gamma_1\sin\theta+(\alpha-\beta)\cos\theta\cos\varphi)^2]
     \nonumber \\
& & + 4\tau_1^2\omega_C\omega_L[(\alpha^2-\beta^2)\cos\theta
    +(\alpha+\beta)\gamma_1\sin\theta \cos\varphi
     +(\beta-\alpha)\gamma_2\sin\theta \sin\varphi]\}
      \label{eq:rate1}
\end{eqnarray}
\begin{eqnarray}
\Gamma_{x'x'}&=&
  \frac{2k^2\tau_1}{\hbar^2}\{\frac{1+(\omega_c^2+\omega_L^2)
    \tau_1^2}{D_+D_-}\times[(\gamma_2\cos\theta+(\alpha+\beta)\sin\theta \sin\varphi)^2
      \nonumber\\
& & +(\gamma_1\cos\theta-(\alpha-\beta)\sin\theta
     \cos\varphi)^2]+\frac{(\alpha-\beta)^2\sin^2\varphi
      +(\alpha+\beta)^2\cos^2\varphi}{1+\omega_C^2\tau_1^2}\}
       \label{eq:rate2}
       \end{eqnarray}
\begin{eqnarray}
\Gamma_{y'y'}&=&\frac{2k^2\tau_1}{\hbar^2}\{\frac{1+(\omega_c^2+\omega_L^2)\tau_1^2}{D_+D_-}
    \times[(\gamma_2\cos\theta+(\alpha+\beta)\sin\theta \sin\varphi)^2
    + (\gamma_1\cos\theta-(\alpha-\beta)\sin\theta \cos\varphi)^2]
    \nonumber\\
& & +\frac{1}{1+\omega_C^2\tau_1^2}\times[(\gamma_2\sin\theta-(\alpha+\beta)\cos\theta
      \sin\varphi)^2+(\gamma_1\sin\theta+(\alpha-\beta)\cos\theta\cos\varphi)^2]\}
     \label{eq:rate3}
\end{eqnarray}
\begin{eqnarray}
\Gamma_{x'y'}&=&\frac{2k^2\tau_1}{\hbar^2}\{\frac{[(\omega_c^2-\omega_L^2)\tau_1^2-1]
  \omega_L\tau_1}{D_+D_-}\times[(\gamma_2\cos\theta+(\alpha+\beta)\sin\theta \sin\varphi)^2
     +(\gamma_1\cos\theta-(\alpha-\beta)\sin\theta\cos\varphi)^2]
     \nonumber\\
& & + \frac{1}{1+\omega_C^2\tau_1^2}\times[-(\alpha^2-\beta^2)\cos\theta\omega_C\tau_1
     +((\alpha-\beta)^2-(\alpha+\beta)^2)\cos\theta\cos\varphi\sin\varphi
     \nonumber\\
& &  +\gamma_1[(\alpha-\beta)\sin\theta\sin\varphi-(\alpha+\beta)\omega_C\tau_1\sin\theta\cos\varphi]
      +\gamma_2[(\alpha-\beta)\omega_C\tau_1\sin\theta\sin\varphi
       +(\alpha+\beta)\sin\theta\cos\varphi]]\}
        \nonumber \\
       \label{eq:rate4}
\end{eqnarray}
\begin{eqnarray}
\Gamma_{y'x'}&=&
 \frac{2k^2\tau_1}{\hbar^2}\{\frac{[1-(\omega_c^2-\omega_L^2)\tau_1^2]
    \omega_L\tau_1}{D_+D_-}\times[(\gamma_2\cos\theta+(\alpha+\beta)\sin\theta \sin\varphi)^2
      +(\gamma_1\cos\theta-(\alpha-\beta)\sin\theta\cos\varphi)^2]
      \nonumber\\
& & + \frac{1}{1+\omega_C^2\tau_1^2}\times[(\alpha^2-\beta^2)\cos\theta\omega_C\tau_1
     + ((\alpha-\beta)^2-(\alpha+\beta)^2)\cos\theta\cos\varphi\sin\varphi
       \nonumber \\
& & + \gamma_1[(\alpha-\beta)\sin\theta\sin\varphi
     +(\alpha+\beta)\omega_C\tau_1\sin\theta\cos\varphi]
       +\gamma_2[(\alpha-\beta)\omega_C\tau_1\sin\theta\sin\varphi
        +(\alpha+\beta)\sin\theta\cos\varphi]]\}
        \nonumber\\
        \label{eq:rate5}
\end{eqnarray}
\end{widetext}
with $k=\sqrt{2mE_F}$ and $D_{\pm}=1+(\omega_L\pm\omega_C)^2\tau_1^2.$

\end{document}